\documentclass[a4paper,10pt,twoside]{cpc-hepnp}
\usepackage{multicol}
\usepackage{graphicx}
\usepackage{booktabs}
\usepackage{amssymb,bm,mathrsfs,bbm,amscd}
\usepackage[tbtags]{amsmath}
\usepackage{lastpage}
\usepackage{CJK}
\usepackage{graphicx}
 \usepackage{float} 
\usepackage[scientific-notation=true]{siunitx}
\usepackage{amsmath}
\usepackage{amsfonts}
\usepackage{amssymb}
\usepackage{multirow}
\usepackage{tikz}
\usetikzlibrary{shapes.geometric, arrows}

\usepackage[utf8]{inputenc}
\usepackage{hyperref}
\def\({\left(}
\def\){\right)}
\def\[{\left[}
\def\]{\right]}

\def\be{\begin{eqnarray}}
\def\ee{\end{eqnarray}}
\usepackage[capitalise]{cleveref}
\crefname{figure}{Fig.}{Figs.}
\Crefname{figure}{Fig.}{Figs.}

\begin{document}
\begin{CJK*}{GBK}{song}

\fancyhead[c]{\small Chinese Physics C~~~Vol. **, No. ** (**)
**} \fancyfoot[C]{\small **-\thepage}

\footnotetext[0]{Received \today}

\title{Impacts of the simulated standard siren on estimating cosmological parameters\thanks{Supported by the National Science Foundation of China under grants No. U1931202, 11633001, and 11690023; the National Key R\&D Program of China No. 2017YFA0402600}}

\author{
  	  Dong Zhao$^{1}$
\quad Zhi-Chao Zhao$^{1}$
\quad Jun-Qing Xia$^{1;1)}$\email{xiajq@bnu.edu.cn}
}
\maketitle

\address{$^1$ Department of Astronomy, Beijing Normal University, Beijing 100875, China\\}

\begin{abstract}
We propose a method based on the process of extracting gravitational wave (GW) parameters from GW signals to simulate the binary neutron-star (BNS) merging events. We simulate 1000 GW standard sirens based on the observation of the Einstein Telescope (ET). Almost all the simulated GW data are in the redshift range of $[0,3]$. The role of the GW standard siren in the inference of the cosmological parameters is investigated. We find that the GW data can help improve the accuracy of cosmological parameters. Moreover, the degeneracy of cosmological parameters is broken by the GW data. The GW standard siren is helpful for the constraint of the cosmological parameters.
\end{abstract}



\footnotetext[0]{\hspace*{-3mm}\raisebox{0.3ex}{$\scriptstyle\copyright$}2020
Chinese Physical Society and the Institute of High Energy Physics
of the Chinese Academy of Sciences and the Institute
of Modern Physics of the Chinese Academy of Sciences and IOP Publishing Ltd}%

\begin{multicols}{2}

\section{Introduction}\label{sec:introduction}\noindent
The cosmological principle implies that the Universe is homogeneous and isotropic on large scales \cite{Weinberg:2008zzc}. Based on it, the $\Lambda$CDM model has been well established. The basic cosmological parameters are constrained to a few percent levels by the most powerful cosmological probes, such as the cosmic microwave background (CMB) anisotropies measurements \cite{Hinshaw:2012aka,Bennett:2012zja,Ade:2013zuv,Ade:2015xua,Aghanim:2018eyx}, the baryon acoustic oscillations (BAO) measurements \cite{Beutler:2011hx,Ross:2014qpa,Cuesta:2015mqa}, and the type Ia supernovae (SN) observations \cite{Riess:1998cb,Perlmutter:1998np,Suzuki:2011hu,Betoule:2014frx}. Recently, a new way of observing luminosity distance independent of cosmological models is provided by the GW observations \cite{TheLIGOScientific:2016wyq,TheLIGOScientific:2017qsa}. It is well-known that the detection of GW170817 is an important milestone in the history of gravitational waves observation. GW170817 is a signal from a merging binary neutron-star observed by the Advanced LIGO and Virgo detectors, in particular, containing GW and corresponding electromagnetic signals \cite{TheLIGOScientific:2017qsa}. The detection of GW170817 means the new era of multi-messenger astronomy is coming.
The GW events that both GW and corresponding electromagnetic signals are observed can be serveed as standard sirens in the study of cosmology \cite{Schutz:1986gp,Holz:2005df}. The result of the Hubble constant given by the GW170817 independently is $H_0=70.0^{+12.0}_{-8.0}$ km $\rm s^{-1}$ $\rm Mpc^{-1}$ \cite{Abbott:2017xzu}. Although the uncertainty is still large, it is undeniable that the GW standard siren is a powerful cosmological probe. 

In the future, people can observe more GW standard sirens through the third-generation ground-based GW detectors, such as the Einstein Telescope \cite{2011einstein}. According to the current design, ET has three detectors with 10 km-long arms, which at an angle of 60 degrees to each other. Compared to the advanced ground-based detectors, ET has ten times more sensitivity in amplitude and it covers a wider detection frequency range of $1-10^{4}$ Hz \cite{2011einstein}. Recently, some works discussed the cosmological parameters constraint based on the observation of the ET in the future, indicating that the GW standard siren observed through ET will be a powerful tool to infer the cosmological parameters space \cite{Cai:2016sby,Zhang:2019ple,Zhang:2018byx,Wang:2018lun,Zhao:2019njc}. When the GW data is combined with other cosmological data such as the CMB, BAO, and SN data in the constraint of cosmological models, the accuracy of the cosmological parameters can be improved and the parameter degeneracies formed by other observations are broken by the GW data \cite{Zhang:2019ple,Zhang:2018byx,Wang:2018lun}.

In this paper, we will propose a method using Bilby \cite{Ashton:2018jfp} to simulate the signals of GW standard sirens observed by ET. The method is based on the process of extracting GW parameters from GW signals. During our work, Bilby is served as a tool to inject signals and infer parameters from the injected signals \cite{Ashton:2018jfp}. The detectors' noise are added to the GW signals by Bilby. Two cosmological models are considered in this work, namely, the $\Lambda$CDM model and the $\omega$CDM model. We get the fiducial parameters by containing the two cosmological models with the current CMB$+$BAO$+$SN (CBS) data. For CMB data, we use the low-$\ell$ and high-$\ell$ Planck temperature power spectra \cite{Aghanim:2019ame}.  For the BAO data, we use the data from 6dFGS \cite{Beutler:2011hx} and SDSS-MGS \cite{Ross:2014qpa}. For the SN data, the JLA compilation is considered \cite{Betoule:2014frx}. The simulated GW data alone will be used to constrain parameters in the two cosmological models. Simultaneously, we will consider a combination of the simulated GW data with the three mainstream cosmological observation data, i.e., CMB, BAO, and SN to constrain the two cosmological models. We will investigate the role of GW standard sirens in the parameter inference process by comparing the results given by different datasets.

The rest of this paper is organized as follows. In Scetion \ref{sec:simulation}, we show the method to simulate the GW data. In Section \ref{sec:CP}, we show the results of constraining cosmological models with different datasets. Discussions and conclusions are given in Section \ref{sec:DC}.

\section{Simulating GW data}\label{sec:simulation}\noindent
The waveform of GWs from the merging BNS can be described by different approximate gravitational waveform models. In our work, we choose the IMRPhenomPv2$\_$NRTidal model \cite{Dietrich:2018uni} to simulate the waveform of GW, which has been used to analyze the long BNS signal GW170817 \cite{Abbott:2018wiz,Abbott:2018exr} and GW190425 \cite{Abbott:2020uma}. 
The redshift distribution of the GW sources observed by the detectors on Earth takes the form \cite{Zhao:2010sz,Cai:2016sby,Zhang:2019ple}
\begin{equation}
P(z) \propto \frac{4 \pi d_{C}^{2}(z) R(z)}{H(z)(1+z)},
\end{equation}
where $H(z)$ is the Hubble parameter. $R(z)$ is the time evolution of the burst rate, which is given as \cite{Cai:2016sby,Wang:2018lun,Schneider:2000sg,Cutler:2009qv}
\begin{equation}
R(z)=\left\{\begin{array}{ll}
1+2 z, & z \leq 1 \\
\frac{3}{4}(5-z), & 1<z<5 \\
0. & z \geq 5
\end{array}\right.
\end{equation}
$d_{C}(z)$ denotes the comoving distance at the redshift $z$, taking the form
\begin{equation}
d_{C}(z)=\frac{1}{H_{0}} \int_{0}^{z} \frac{d z^{\prime}}{E\left(z^{\prime}\right)},
\end{equation}
where $E(z)=H(z) / H_{0}$ is the dimensionless Hubble parameter. The expression of $E(z)$ depends on the cosmological models. In the $\Lambda$CDM model, $E(z)$ takes the form
\begin{equation}
E^{2}(z)=\Omega_{m}(1+z)^{3}+\left(1-\Omega_{m}\right),
\end{equation}
where $\Omega_{m}$ is the matter density at the present epoch. In the $\omega$CDM model, $E(z)$ has the form
\begin{equation}
E^{2}(z)=\Omega_{m}(1+z)^{3}+\left(1-\Omega_{m}\right)(1+z)^{3\left(1+\omega\right)},
\end{equation}
where $\omega=p/\rho$ denotes the equation of state of dark energy.

In the simulation, we choose the mass of the two neutron stars are both 1.4 $M_{\odot}$, which is the typical mass of neutron star \cite{seeds2009astronomy}. The neutron stars in a binary system are usually considered as old stars, so the effect from the spins of neutron stars can be ignored. The tidal deformability parameters that have little effect on the results are both set as 425 \cite{TheLIGOScientific:2017qsa}. We assume the distribution of GW events are homogeneously on the celestial sphere. After generating the waveform, we can then calculate the combined optical SNR for the network of three independent interferometers. It can be written as
\begin{equation}
\rho=\sqrt{\sum_{i=1}^{3}\left(\rho^{(i)}\right)^{2}},
\end{equation}
where $\rho^{(i)}=\sqrt{\left\langle\mathcal{H}^{(i)}, \mathcal{H}^{(i)}\right\rangle}$ and the inner product can be calculated by
\begin{equation}
\langle a, b\rangle=4 \int_{f_{\text {lower }}}^{f_{\text {upper }}} \frac{\tilde{a}(f) \tilde{b}^{*}(f)+\tilde{a}^{*}(f) \tilde{b}(f)}{2} \frac{d f}{S_{h}(f)},
\end{equation}
where $\tilde{a}(f)$ and $\tilde{b}(f)$ denote the Fourier transforms of the functions $a(f)$ and $b(f)$, respectively. $S_{h}(f)$, the one-side noise power spectral density (PSD), is used to describe the performance of a GW detector. The $f_{\text {lower }}$ is 1 Hz and the $f_{\mathrm{LSO}}=1 /\left(6^{3 / 2} 2 \pi M_{\mathrm{obs}}\right)$, where the $M_{\mathrm{obs}}$ is the observed total mass. The power spectral density used in this work is given in Ref. \cite{Zhao:2010sz}. Only if the optimal SNR of injection is larger than 8, we think it is detectable. We drop the injection whose optimal SNR $<$ 8. 

The results of the best-fitted $\Lambda$CDM and $\omega$CDM model, which constrained by the current CBS data, serve as the fiducial parameters for the two cosmology models. We list the fiducial parameters in Table \ref{table:fucial}.
\begin{table}[H]
	\begin{center}
	\caption{The fiducial parameters of the best-fitted $\Lambda$CDM and $\omega$CDM model with the current CBS data.}
	\setlength{\tabcolsep}{3mm}{
		\begin{tabular}{ccccccc}
			\hline Model &  $\Lambda$CDM && $w$CDM & \\
			\hline$\Omega_{\mathrm{m}}$ & $0.304$ && $0.305$ \\
			$H_{0}$ & $68.10$ && $67.92$ \\
			$\omega$ & $-$ && $-0.99$ \\
			\hline
	\end{tabular}}
	\label{table:fucial}
	\end{center}
\end{table}

\begin{figure}[H]
	\centering
		\includegraphics[width=8cm]{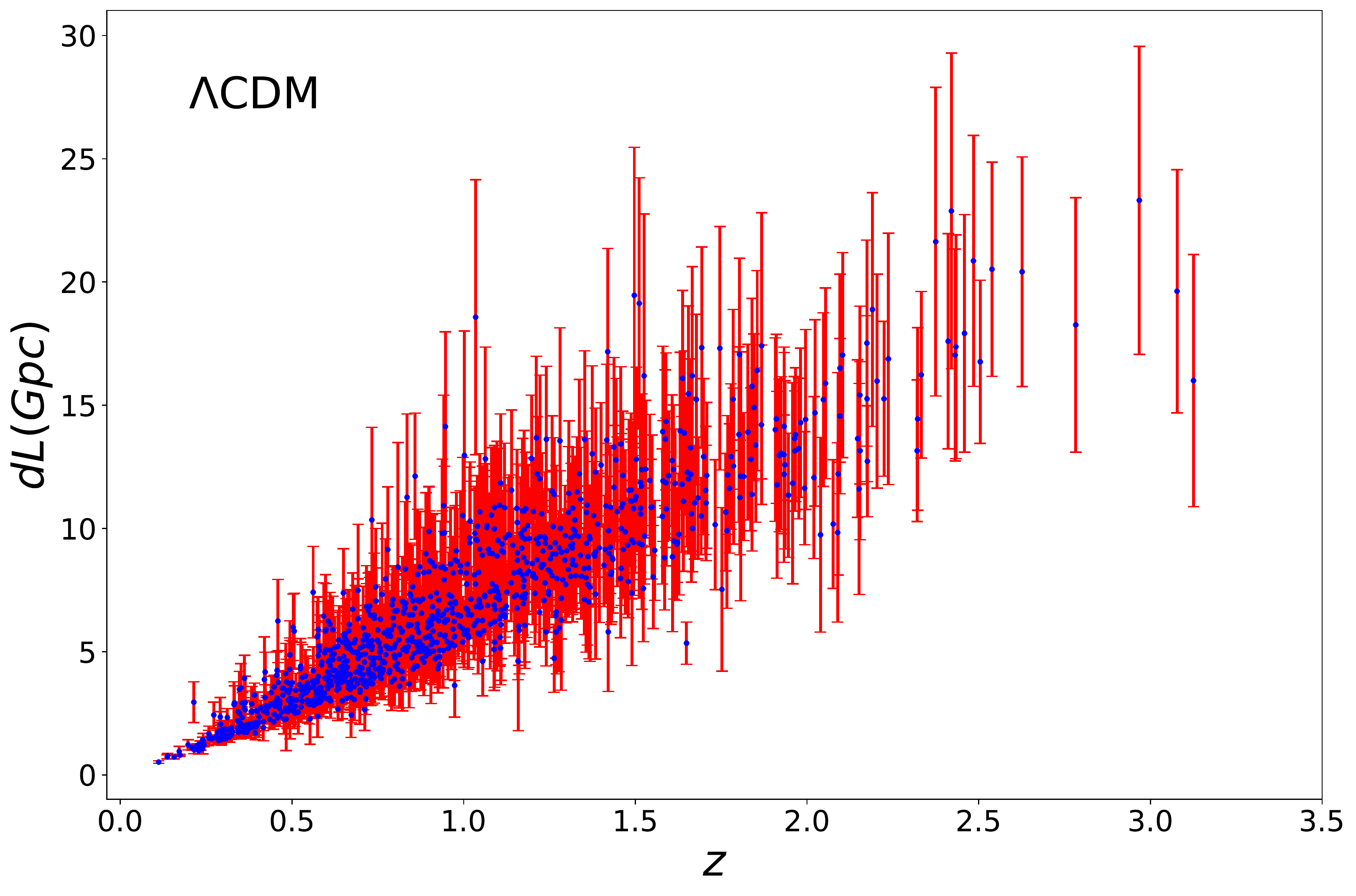}
		\caption{The luminosity distance and its uncertainty inferred from 1000 simulated GW events based on the best-fitted $\Lambda$CDM model. The fiducial parameters are derived from the current CBS data.}
		\label{fig:LCDM_dl}
\end{figure}

\begin{figure}[H]
	\centering
		\includegraphics[width=\columnwidth]{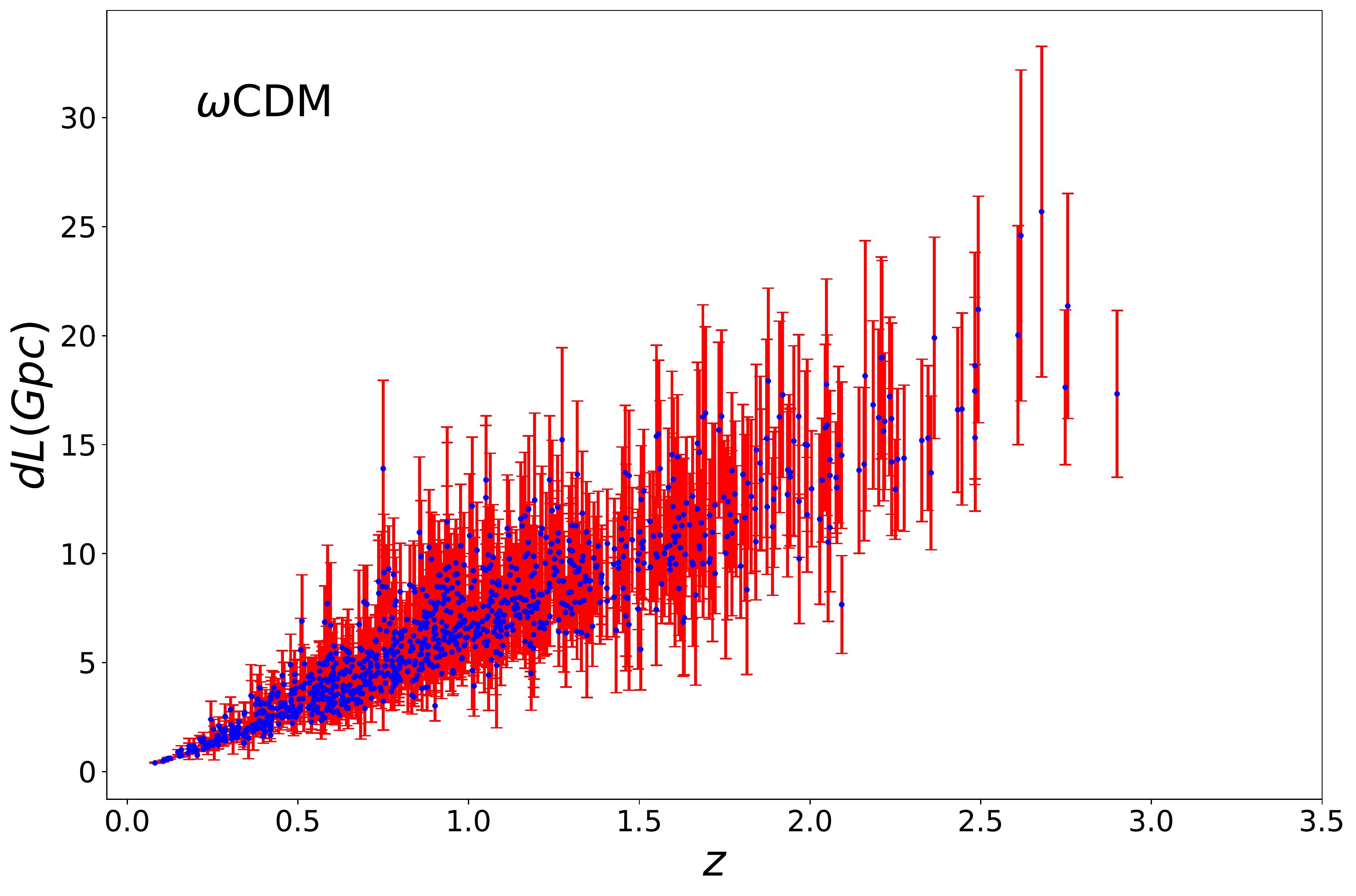}
		\caption{The luminosity distance and its uncertainty inferred from 1000 simulated GW events based on the best-fitted $\omega$CDM model. The fiducial parameters are derived from the current CBS data.}
		\label{fig:wCDM_dl}
\end{figure}

We simulate 1000 BNS merging events based on the observation of ET. The luminosity distance inferred from the simulated GWs data has different upper and lower limit of deviation. For convenience, we consider an approach to combine them, namely,
\begin{equation}
\sigma_{d_{L}}=\sqrt{\frac{\left(\sigma_{d_{L}}^{+}\right)^{2}+\left(\sigma_{d_{L}}^{-}\right)^{2}}{2}}.
\end{equation}
The luminosity distance and its uncertainty inferred from the simulated GWs data are shown in Fig. \ref{fig:LCDM_dl}-\ref{fig:wCDM_dl}. The Fig. \ref{fig:LCDM_dl} denotes the result based on the best-fitted $\Lambda$CDM model and the Fig. \ref{fig:wCDM_dl} denotes the result related to the best-fitted $\omega$CDM model. We can see from Fig. \ref{fig:LCDM_dl}-\ref{fig:wCDM_dl} that the maximum redshift of the simulated GW sources are around 3. The maximum redshift of the simulated GW sources in Fig. \ref{fig:LCDM_dl} is 3.126 and the one in FIg. \ref{fig:wCDM_dl} is 2.899. There are only a few GW sources that have the redshift large than 3.

\section{Constraing cosmological parameters with the simulated GW data}\label{sec:CP}\noindent
In our work, we use the Markov-chain Monte Carlo (MCMC) \cite{Lewis:2002ah} method to explore the cosmological parameters space. For the 1000 simulated GW data, we employ the $\chi^{2}$ function as follow
\begin{equation}
\chi_{\mathrm{GW}}^{2}=\sum_{i=1}^{1000}\left[\frac{d_{L}^{i}-d^{'}_{L}\left(z_{i} ; \vec{\Omega}\right)}{\sigma_{d_{L}}^{i}}\right]^{2},
\end{equation}
where the $z_{i}$, $d_{L}^{i}$ and $\sigma_{d_{L}}^{i}$ indicates the $i$th redshift, luminosity distance, and error of luminosity distance, respectively. $d^{'}_{L}\left(z_{i} ; \vec{\Omega}\right)$ is the luminosity distance given by the cosmological model. $\vec{\Omega}$ represents the collection of the cosmological parameters. 

\begin{figure*}
	\begin{center}
		\includegraphics[width=12cm]{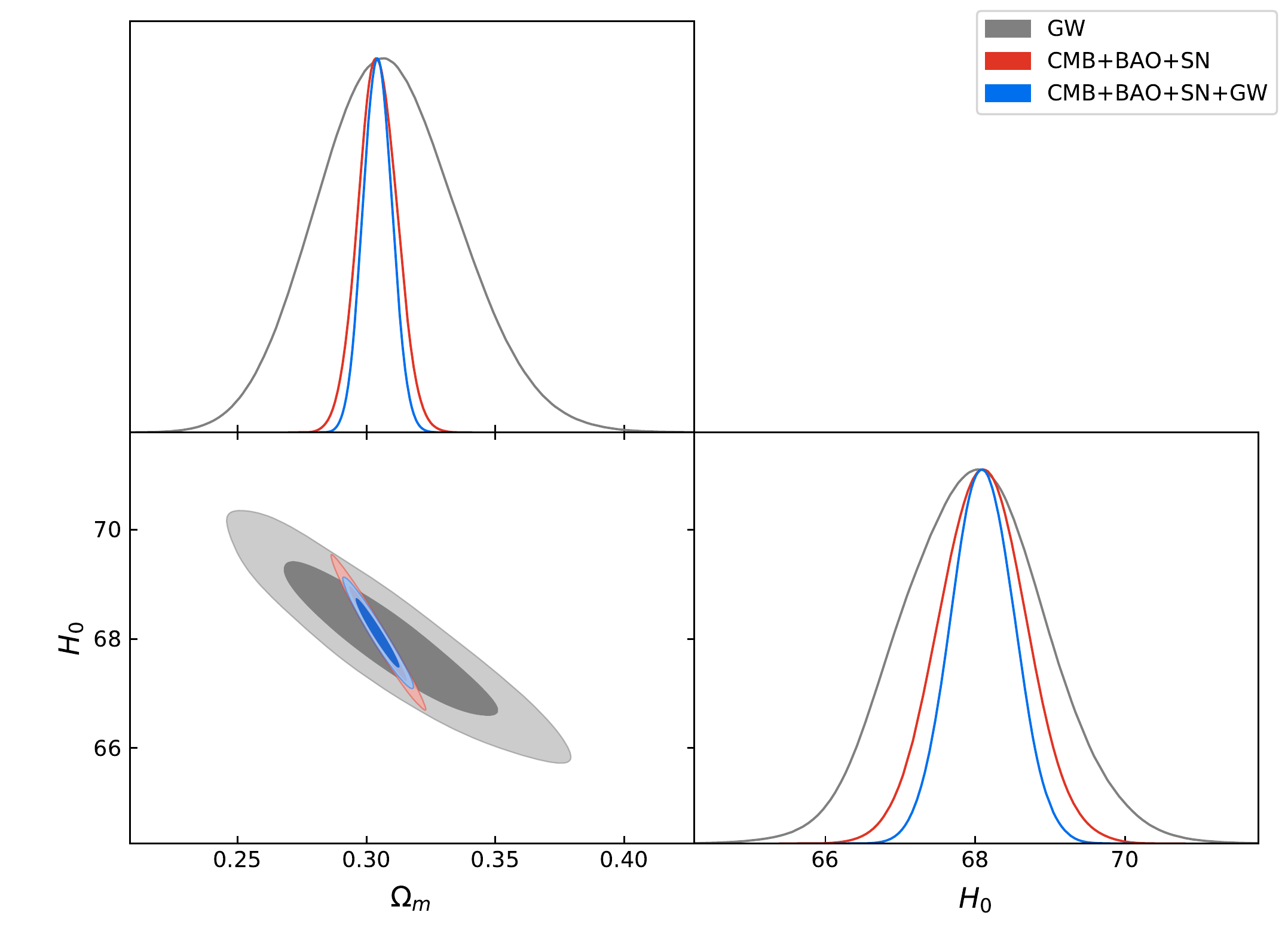}
		\caption{The 1-dimensional and 2-dimensional marginalized posterior distributions of the parameters $H_0$ and $\Omega_{\mathrm{m}}$ in the $\Lambda$CDM model, which given by the GW, CBS and, CBS+GW data. The colors, i.e., grey, red, and, blue denote the results related to the GW, CBS, and CBS+GW data, respectively.}
		\label{fig:LCDM_C}
	\end{center}
\end{figure*}

\begin{figure*}
	\begin{center}
		\includegraphics[width=12cm]{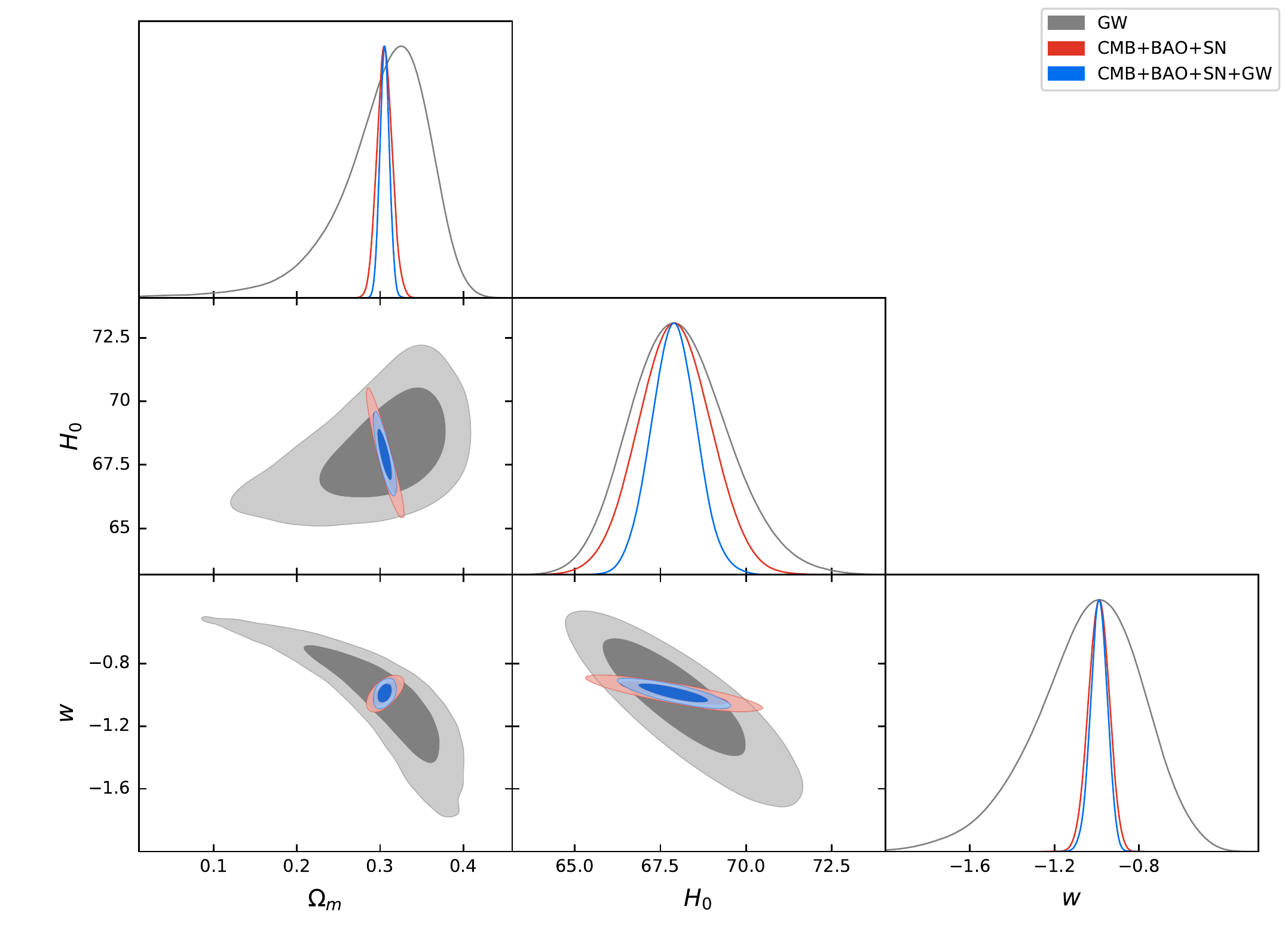}
		\caption{The 1-dimensional and 2-dimensional marginalized posterior distributions of the parameters $H_0$, $\Omega_{\mathrm{m}}$ and $\omega$ in the $\omega$CDM model, which given by the GW, CBS and, CBS+GW data. The colors, i.e., grey, red, and, blue denote the results related to the GW, CBS, and, CBS+GW data, respectively.}
		\label{fig:wCDM_C}
	\end{center}
\end{figure*}

We use three datasets to constrain the $\Lambda$CDM and $\omega$CDM model, namely the GW, CBS, and the combination of CBS+GW data. The 1-dimensional and 2-dimensional marginalized posterior distributions of the parameters are shown in Fig. \ref{fig:LCDM_C}-\ref{fig:wCDM_C} and the colors, i.e., grey, red, and, blue denote the results given by the GW, CBS, and CBS+GW data, respectively. 
The 68.3\% confidence level constraints on the cosmological parameters are summarized in Table \ref{table:data}. Overall, the simulated GW data whose redshifts are almost in the range of $[0,3]$ has a weak constraint on the cosmological parameters. However, when we combine the GW data with other observation data, we find that the GW data can improve the constraint accuracy of cosmological paramters. Moreover, the degeneracy orientations of the GW and CBS data are different in the parameter planes, especially for the $\omega$CDM model. Thus, the GW data can break the degeneracies between cosmological parameters. 
\begin{table*}
	\centering
	\caption{The best fitting values for the $\Lambda$CDM and $\omega$CDM model constrained by the GW, CBS, and, CBS+GW data. We show the 68.3\% confidence level constraints on the parameters $\Omega_{\mathrm{m}}$, $H_{0}$, and $\omega$.}
	\setlength{\tabcolsep}{1.8mm}{
		\begin{tabular}{ccccccc}
			\hline Model & & $\Lambda$CDM & & & $w$CDM & \\
			\hline Data & GW & CBS & CBS+GW & GW & CBS & CBS+GW \\
			\hline$\Omega_{\mathrm{m}}$ & $0.309_{-0.0286}^{+0.0256}$ & $0.304_{-0.0073}^{+0.0075}$ & $0.304 \pm 0.0056$ & $0.302_{-0.0326}^{+0.0644}$ & $0.305_{-0.0091}^{+0.0090}$ & $0.305 \pm 0.0056$ \\
			$H_{0}$ & $67.98_{-0.968}^{+0.936}$ & $68.10_{-0.576}^{+0.571}$ & $68.10_{-0.417}^{+0.423}$ & $68.08_{-1.538}^{+1.257}$ & $67.92_{-1.048}^{+1.047}$ & $67.90_{-0.657}^{+0.655}$ \\
			$\omega$ & $-$ & $-$ & $-$ & $-1.042_{-0.2075}^{+0.2866}$ & $-0.990_{-0.0482}^{+0.0483}$ & $-0.989 \pm 0.0396$ \\
			\hline
	\end{tabular}}
	\label{table:data}
\end{table*}

The results of constraint on the $\Lambda$CDM model are shown in Fig. \ref{fig:LCDM_C} and summarized in Table \ref{table:data}. The measurements given by GW data alone are $1.4\%$ and $8.78\%$ for parameters $H_0$ and $\Omega_{\mathrm{m}}$, respectively. For the CBS data, the accuracies of $H_0$ and $\Omega_{\mathrm{m}}$ are $0.84\%$ and $2.43\%$, respectively. When we combinate the GW and CBS data, the accuracy of $H_0$ is improved to $0.62\%$ and the accuracy of $\Omega_{\mathrm{m}}$ is improved to $1.84\%$. It shows that the GW data can improve the accuracies of $H_0$ and $\Omega_{\mathrm{m}}$ in the $\Lambda$CDM model. Besides, the degeneracy orientations given by the GW and CBS data in the $\Omega_{\mathrm{m}}-H_{0}$ plane are different so that the parameter degeneracy of the two parameters is broken by the GW data.

The results of constraint on the $\omega$CDM model are shown in Fig. \ref{fig:wCDM_C} and summarized in Table \ref{table:data}. Compared with the $\Lambda$CDM model, the GW data alone provides a much worse constraint on the $\omega$CDM model. For the GW data alone, the accuracies of parameters $H_0$, $\Omega_{\mathrm{m}}$, and $\omega$ are $2.06\%$, $16.9\%$ and $24.01\%$, respectively. Similarly with the $\Lambda$CDM model, when we combinate the GW and CBS data, the accuracies of parameters are improved. The accuracy of parameter $H_0$ is $1.54\%$ for CBS data and it is improved to $0.97\%$ when the GW data is considered in the fitting. For the parameter $\Omega_{\mathrm{m}}$, the accuracy is improved from $2.97\%$ to $1.24\%$ by the GW data. The GW data is not very helpful in the constraint of the parameter $\omega$ so that the accuracy of $\omega$ is improved slightly from $4.87\%$ to $4\%$. In Fig. \ref{fig:wCDM_C}, we can see that there is an angle between the degeneracy orientations given by the CBS and GW data in the $H_{0}-\omega$ plane so that the parameter degeneracy of $H_0$ and $\omega$ is broken by the GW data. More particularly, the degeneracy orientations are almost orthogonal in the $\Omega_{\mathrm{m}}-H_{0}$ and $\Omega_{\mathrm{m}}-\omega$ planes, indicating that the parameter degeneracies of $\Omega_{\mathrm{m}}-H_{0}$ and $\Omega_{\mathrm{m}}-\omega$ are thoroughly broken by the GW data.

\section{Discussions and conclusions}\label{sec:DC}\noindent
\vspace{0.1cm}
In this paper, we proposed a method based on the process of extracting GW parameters from GW signals to simulate the BNS merging events.
We simulated 1000 BNS merging events based on the observation of the Einstein Telescope. Almost all the redshifts of the simulated GW data are in the range of $[0, 3]$. We considered two cosmological models, namely, the $\Lambda$CDM model and the $\omega$CDM model. The fiducial parameters are given by the combination of CMB, BAO, and, SN data. The role of the GW standard sirens in the cosmological parameters measurement was investigated. We found that the accuracies of cosmological parameters are improved when the GW data are considered in the fitting. In the $\Lambda$CDM model, the accuracy of the parameter $\Omega_{\mathrm{m}}$ is $2.43\%$ with CBS data and it is improved to $1.84\%$ by the GW data. For the parameter $H_0$, the accuracy is improved from $0.84\%$ to $0.62\%$ by the GW data. In the $\omega$CDM model, the CBS+GW data can provide a $1.24\%$ measurement for the parameter $\Omega_{\mathrm{m}}$, much better than a $2.97\%$ measurement which given by the CBS data. For the parameter $H_0$,  the accuracy is improved from $2.97\%$ to $1.24\%$ by the GW data. The GW data is not very helpful in the constraint of the parameter $\omega$ so that the accuracy of $\omega$ is improved slightly from $4.87\%$ to $4\%$. We also found that the GW data can break the degeneracy of the cosmological parameters, especially for the $\omega$CDM model. In the $\Lambda$CDM model, the degeneracy orientations in the $\Omega_{\mathrm{m}}-H_{0}$ plane given by the GW and CBS data are different, resulting in the breaking of the parameter degeneracy. In the $\omega$CDM model, the parameter degeneracy of $H_0$ and $\omega$ is also broken by the GW data. More particularly, the degeneracy orientations are almost orthogonal in the $\Omega_{\mathrm{m}}-H_{0}$ and $\Omega_{\mathrm{m}}-\omega$ planes, indicating that the parameter degeneracies of $\Omega_{\mathrm{m}}-H_{0}$ and $\Omega_{\mathrm{m}}-\omega$ are thoroughly broken by the GW data. Our results show that the GW standard sirens can be very helpful in the inference of the cosmological parameters.

\acknowledgments{We thank Yong Zhou for helpful discussions.}

\vspace{10mm}

\vspace{-1mm}
\centerline{\rule{80mm}{0.1pt}}

\end{multicols}

\clearpage

\end{CJK*}

\vspace{2mm}


\begin{thebibliography}{10}
	\expandafter\ifx\csname url\endcsname\relax
	\def\url#1{{\tt #1}}\fi
	\expandafter\ifx\csname urlprefix\endcsname\relax\def\urlprefix{URL }\fi
	\providecommand{\eprint}[2][]{\url{#2}}
	
\bibitem{Weinberg:2008zzc}
S. Weinberg, {\em {Cosmology}} (PUBLISHER, ADDRESS, 2008).

\bibitem{Hinshaw:2012aka}
G. Hinshaw {\it et~al.}, Astrophys. J. Suppl. {\bf 208},  19  (2013).

\bibitem{Bennett:2012zja}
C. Bennett {\it et~al.}, Astrophys. J. Suppl. {\bf 208},  20  (2013).

\bibitem{Ade:2013zuv}
P. Ade {\it et~al.}, Astron. Astrophys. {\bf 571},  A16  (2014).

\bibitem{Ade:2015xua}
P. Ade {\it et~al.}, Astron. Astrophys. {\bf 594},  A13  (2016).

\bibitem{Aghanim:2018eyx}
N. Aghanim {\it et~al.},   (2018).

\bibitem{Beutler:2011hx}
F. Beutler {\it et~al.}, Mon. Not. Roy. Astron. Soc. {\bf 416},  3017  (2011).

\bibitem{Ross:2014qpa}
A.~J. Ross {\it et~al.}, Mon. Not. Roy. Astron. Soc. {\bf 449},  835  (2015).

\bibitem{Cuesta:2015mqa}
A.~J. Cuesta {\it et~al.}, Mon. Not. Roy. Astron. Soc. {\bf 457},  1770
(2016).

\bibitem{Riess:1998cb}
A.~G. Riess {\it et~al.}, Astron. J. {\bf 116},  1009  (1998).

\bibitem{Perlmutter:1998np}
S. Perlmutter {\it et~al.}, Astrophys. J. {\bf 517},  565  (1999).

\bibitem{Suzuki:2011hu}
N. Suzuki {\it et~al.}, Astrophys. J. {\bf 746},  85  (2012).

\bibitem{Betoule:2014frx}
M. Betoule {\it et~al.}, Astron. Astrophys. {\bf 568},  A22  (2014).

\bibitem{TheLIGOScientific:2016wyq}
B. Abbott {\it et~al.}, Phys. Rev. Lett. {\bf 116},  131102  (2016).

\bibitem{TheLIGOScientific:2017qsa}
B. Abbott {\it et~al.}, Phys. Rev. Lett. {\bf 119},  161101  (2017).

\bibitem{Schutz:1986gp}
B.~F. Schutz, Nature {\bf 323},  310  (1986).

\bibitem{Holz:2005df}
D.~E. Holz and S.~A. Hughes, Astrophys. J. {\bf 629},  15  (2005).

\bibitem{Abbott:2017xzu}
B. Abbott {\it et~al.}, Nature {\bf 551},  85  (2017).

\bibitem{2011einstein}
{\em Einstein Gravitational Wave Telescope Conceptual Design Study:
	ET-0106C-10} (European gravitational observatory, ADDRESS, 2011).

\bibitem{Cai:2016sby}
R.-G. Cai and T. Yang, Phys. Rev. D {\bf 95},  044024  (2017).

\bibitem{Zhang:2019ple}
J.-F. Zhang, H.-Y. Dong, J.-Z. Qi, and X. Zhang, Eur. Phys. J. C {\bf 80},  217
(2020).

\bibitem{Zhang:2018byx}
X.-N. Zhang, L.-F. Wang, J.-F. Zhang, and X. Zhang, Phys. Rev. D {\bf 99},
063510  (2019).

\bibitem{Wang:2018lun}
L.-F. Wang, X.-N. Zhang, J.-F. Zhang, and X. Zhang, Phys. Lett. B {\bf 782},
87  (2018).

\bibitem{Zhao:2019njc}
Z.-C. Zhao, H.-N. Lin, and Z. Chang, Chin. Phys. C {\bf 43},  075102  (2019).

\bibitem{Ashton:2018jfp}
G. Ashton {\it et~al.}, Astrophys. J. Suppl. {\bf 241},  27  (2019).

\bibitem{Aghanim:2019ame}
N. Aghanim {\it et~al.},   (2019).

\bibitem{Dietrich:2018uni}
T. Dietrich {\it et~al.}, Phys. Rev. D {\bf 99},  024029  (2019).

\bibitem{Abbott:2018wiz}
B. Abbott {\it et~al.}, Phys. Rev. X {\bf 9},  011001  (2019).

\bibitem{Abbott:2018exr}
B. Abbott {\it et~al.}, Phys. Rev. Lett. {\bf 121},  161101  (2018).

\bibitem{Abbott:2020uma}
B. Abbott {\it et~al.}, Astrophys. J. Lett. {\bf 892},  L3  (2020).

\bibitem{Zhao:2010sz}
W. Zhao, C. Van Den~Broeck, D. Baskaran, and T. Li, Phys. Rev. D {\bf 83},
023005  (2011).

\bibitem{Schneider:2000sg}
R. Schneider, V. Ferrari, S. Matarrese, and S.~F. Portegies~Zwart, Mon. Not.
Roy. Astron. Soc. {\bf 324},  797  (2001).

\bibitem{Cutler:2009qv}
C. Cutler and D.~E. Holz, Phys. Rev. D {\bf 80},  104009  (2009).

\bibitem{seeds2009astronomy}
M. Seeds and D. Backman, {\em Astronomy: The Solar System and Beyond} (Cengage
Learning, ADDRESS, 2009).

\bibitem{Lewis:2002ah}
A. Lewis and S. Bridle, Phys. Rev. D {\bf 66},  103511  (2002).


\end{thebibliography}
\end{document}